# Comparison of feature extraction tools for network traffic data

Borys Lypa[1,*,†], Ivan Horyn[1,†], Natalia Zagorodna[1,†], Dmytro Tymoshchuk[1] and Taras Lechachenko[1]

[1] *Ternopil Ivan Puluj National Technical University, Ruska str., 56, Ternopil, Ukraine*

**Abstract**
The comparison analysis of the most popular tools to extract features from network traffic is conducted in this paper. Feature extraction plays a crucial role in Intrusion Detection Systems (IDS) because it helps to transform huge raw network data into meaningful and manageable features for analysis and detection of malicious activities. The good choice of feature extraction tool is an essential step in construction of Artificial Intelligence-based Intrusion Detection Systems (AI-IDS), which can help to enhance the efficiency, accuracy, and scalability of such systems.

## 1. Introduction

With ever-increasing amounts of data passed through the internet, the problem of analyzing network traffic has gotten more complex and challenging than ever before. Network traffic analysis is the process of monitoring and analyzing network data flows to gain insights into the performance, security, and management of a computer network. Network traffic analysis involves packet inspection, traffic classification, anomaly detection, security monitoring, performance optimization. Packet inspection and traffic classification can be useful for identifying and blocking malicious traffic, or for prioritizing traffic for different applications. Anomaly detection involves detecting unusual patterns in network traffic, which can be useful for identifying malicious activity, such as denial-of-service attacks or intrusions. Performance monitoring involves monitoring network performance to identify bottlenecks and other problems.

This paper mostly describes a perspective of using network traffic analysis for network security in Intrusion Detection Systems (IDS).

---

One of the most popular IDSs are signature-based IDS that can detect attack/intrusion based on its "signature" by comparing data traffic against known attack signatures available in the relevant database [1]. These signatures are essentially patterns or characteristics associated with known threats or attacks. When the IDS detects a match between the incoming data and any signature in its database, it raises an alert or takes predefined actions to mitigate the threat. [2] This approach can give accurate detection of known attacks [3].

Nowadays the threat landscape is becoming more and more sophisticated and dynamic, requiring constant vigilance and innovative security measures to stay ahead of cyber adversaries. Signature-based network Intrusion Detection Systems are becoming less capable of detecting new threats. That's why cybersecurity companies all around the world invest huge amounts of resources into developing a reliable approach for the detection of network threats using AI algorithms. AI-based IDS show excellent results at anomaly detection, identification of activities or behaviors that significantly deviate from normal patterns. The need for AI-based systems is only increasing [4].

AI-based IDS encompasses machine learning techniques. Data collection and feature space construction is an essential and important step in development and analysis of Machine Learning-Based IDS. Data quality is as important as the choice of algorithm for such systems. Machine learning algorithms require input to be organized in a feature space which can be considered as a dataset of objects with relevant feature values. Network traffic is presented in a format of raw packet capture which could not be used for Machine Learning algorithms in its original form. Feature extraction is a process of transforming raw network traffic data into a set of features that represent characteristics of the traffic such as flow, packet, statistical, time-based, and frequency-based features. Feature selection can contribute to dimensionality and noise reduction, increasing efficiency and accuracy of IDS. But feature extraction is not a strictly predefined process. Different sets of features can be generated from the same raw network traffic data and they can contribute differently in final model efficiency.

A lot of researchers [5, 6, 7] use open datasets with extracted in advance features in order to construct and analyze their models. But such models cannot be easily embedded in IDS because they cannot work with original raw network traffic data. Moreover, we cannot really prognose the accuracy of such models based on real data.

## 2. Introduction to Network traffic data

In previous years, network administrators typically monitored a limited number of devices, usually operated with less than a thousand computers and network bandwidth often restricted to 100 Mbps. Nowadays , administrators contend with high-speed wired networks exceeding 1 Gbps, along with a diverse variety of wireless networks. That's why administrators have to rely on advanced traffic analysis tools in order to effectively manage networks, promptly address issues, prevent failures, and ensure security. Despite the fact that network traffic analysis facilitates robust security management, several challenges have recently arisen. Analysis has to be conducted across multiple levels, including packet, flow, and network levels. Researchers employ various techniques within a generic framework for network traffic analysis, involving preprocessing, subsequent analysis, and observation to discover patterns from network data.

Analyzing network data can be considered as a big data problem due to several factors. Firstly, the total volume of data generated by network devices, such as routers, switches, and servers, is immense. These devices continuously produce logs, packets, and other forms of data that need to be processed and analyzed. With the proliferation of internet-connected devices and the growth of digital communication, this volume is only increasing. It is estimated that the average person produces 1.7 MB per second or 6,120 MB per hour. The average number of members globally is 3.45 on a household scale, meaning a family can create about 506,736 MB daily. [8]

Secondly, network data is often generated at high velocity. Data packets are transmitted rapidly across networks, and real-time analysis is often necessary to detect and respond to security threats, performance issues, or anomalies.

Thirdly, network data encompasses various types of data, including packet headers, payload content, session logs, flow records, and more. Analyzing this diverse range of data sources requires flexible and scalable processing techniques.

Overall, the combination of volume, velocity and variety makes network data a prime candidate for big data analytics. It requires scalable infrastructure, sophisticated algorithms, and efficient processing mechanisms to derive actionable insights from the huge amount of data generated by modern networks.

## 3. Feature extraction for network traffic data

Various techniques can be employed to derive characteristics from a network connection. The most commonly used methods encompass:
- Packet capture and analysis: it involves capturing packets of network traffic and analyzing them to extract features such as packet size, protocol, source and destination IP addresses, port numbers, and flags.
- Flow analysis: it involves grouping packets into flows, which are sequences of packets of the same communication session. Flow size, duration, and protocol are flow features that can be extracted from network traffic flow.
- Application layer monitoring: it involves monitoring network traffic at the application layer to extract features such as the type of application traffic, the URLs accessed, and the amount of data transferred.

Features are often extracted from Packet captures or PCAP. It's a file format used to store network traffic captured by packet sniffers or network monitoring tools. These tools capture data packets as they traverse a network, including such information as source and destination IP addresses, ports, protocols, and the contents of the packets themselves. PCAP are commonly used files for network analysis, troubleshooting, and security purposes, which allow cybersecurity researchers to inspect network traffic for anomalies, malicious activity, or performance issues.

The specific features that are extracted will depend on the specific application and will affect algorithm efficiency. For example, an application, used to detect malicious traffic, may extract different features compared to an application, used to monitor network performance.

## 3.1. Packet level features

Packet-based Intrusion Detection Systems (IDS) are highly regarded for their flexibility in intrusion detection patterns due to the comprehensive volume of data they capture, including all headers up to the application layer (OSI layer 7) and the complete payload. This enables precise rule definition on any part of the traffic, resulting in less number of false positives and higher alert confidence. Implementing simple packet-based IDS is comparatively straightforward, as there's no need to decode protocols beforehand. In some scenarios, when complete data is available in real-time, packet-based IDSs do an excellent job with maximum time resolution. However, encrypted payloads pose a challenge for such systems as signature matching becomes impractical and decreases detection performance.

Packet-based IDSs process all traffic forwarded to them, which lead to higher resource usage. Filtering, aggregation, and state handling are managed entirely on the IDS machine: either through libraries like PCAP or within the IDS itself. Packet-based IDS, particularly without hardware pre-filters, can have a significant processing load, which can lead to system overload. Additionally, packet-based IDSs receive the full payload data of every packet, raising issues about the exposure of confidential information. [9]

Packet-level features include, but not limited to packet size, protocol, source, destination, IP addresses, port numbers, flags.

## 3.2. Flow level features

Cisco NetFlow is a proprietary but openly documented format for transmitting aggregated network data, widely recognized as a standard for flow records. Although it is primarily used for network monitoring rather than intrusion detection, its widespread implementation allows it to be further exploited for intrusion detection without additional computational expenses. Despite its original purpose was not intrusion detection, research demonstrates its effectiveness in detecting certain attacks.

Flow-based feature extraction often relies on the NetFlow protocol. NetFlow records typically contain aggregated data up to the network layer (OSI layer 3), and, depending on probe configuration, may include specific transport (OSI layer 4) information such as TCP flags. Due to the restricted data available in flow-based IDS, defining precise detection rules may not always be feasible, potentially leading to reduced alert confidence and increased false positives.

The generation of flow records introduces a delay between connection establishment and record transmission to the IDS. Depending on configuration, records may only be emitted post-connection closure or timeout, potentially affecting time-sensitive intrusion detection tasks.

Flows may not have sufficient time resolution for some intrusion detection needs, such as determining byte transmission timings. However, encrypted payloads do not interfere with flow-based IDS functionality.

NetFlow data, being aggregated, results in reduced processing requirements for the IDS, generally lowering resource usage. Additionally, NetFlow data poses fewer privacy concerns because most of the potentially sensitive content of the connection remains within the transmission network.

Some of the flow features could be extracted are flow size, flow duration, protocol, source and destination IP addresses.

# 4. Most popular tools

## 4.1. CICflowmeter

CICFlowMeter is a network traffic flow generator and analyzer. [5,6] It can be used to generate bidirectional flows, where the first packet determines the forward (source to destination) and backward (destination to source) directions, hence more than 80 statistical network traffic features such as Duration, Number of packets, Number of bytes, Length of packets, etc. can be calculated separately in the forward and backward directions.

Additional functionalities include, selecting features from the list of existing features, adding new features, and controlling the duration of flow timeout. The output of the application is the CSV format file that has six columns labeled for each flow (FlowID, SourceIP, DestinationIP, SourcePort, DestinationPort, and Protocol) with more than 80 network traffic analysis features.

CICFlowMeter was used for the creation of CIC-IDS2017 Intrusion detection evaluation dataset. [10, 12] And was used for similar network threat detection research, such as [11].

The original tool is available as Java package and source code [13]. There is also Python version created by the community [14].

## 4.2. Wireshark

Wireshark is a widely used network protocol analyzer. It lets you capture and interactively browse the traffic running on a computer network in real-time. It's available for various platforms like Windows, macOS, and Linux. [15]

Wireshark can capture data from a live network connection or read data from a file. It supports hundreds of protocols and can display the captured data in a user-friendly format. This makes it an invaluable tool for network troubleshooting, analysis, software and protocol development, and education.

It provides detailed information about network packets, including their source and destination addresses, protocols used, packet size, and even the contents of individual packets. This level of insight into network traffic is crucial for diagnosing network issues, detecting security breaches, and understanding how applications communicate over a network.

## 4.3. Argus

Argus is a network flow monitoring tool used for collecting and analyzing network traffic data. It differs from packet sniffers like Wireshark in that it focuses on summarizing network flows rather than capturing and analyzing individual packets.

Argus monitors network traffic and generates flow records containing information such as source and destination IP addresses, port numbers, protocol types, timestamps, and packet counts. These flow records provide a higher-level view of network activity, making it easier to identify trends, detect anomalies, and analyze network performance.

One of the key features of Argus is its ability to generate and export flow records in various formats, such as ASCII, binary, and XML. This flexibility allows for seamless integration with other network monitoring and analysis tools. [16]

## 4.4. Snort

Many Network Intrusion Detection Systems (NIDS), whether they rely on misuse or anomaly detection methods, typically operate at the packet level. Among the most widely used open-source tools for intrusion detection is Snort, known for its simplicity and lightweight design, aligning with the misuse detection approach. Essentially, users input signatures, either in string format or more advanced regular expressions, into the system. Its compact size facilitates swift deployment across various network nodes. The system then scrutinizes network traffic for matches with the provided signatures, triggering alarms upon detection. While effective for a limited number of signatures, this approach often falters when faced with a large number of signatures or heavy traffic volumes. Similarly, other systems employing signature-based intrusion detection encounter comparable challenges.

It has a community version available for free, the subscription includes the latest up-to-date rules.

## 4.5. Zeek

Zeek, an open-source network intrusion detection system, while less popular than Snort, boasts compatibility with various Unix flavors. Its architecture is built on multiple layers, spanning from traffic capture to in-depth analysis, enabling easy extension of its capabilities. Zeek offers the flexibility to integrate Snort rules into its framework. Although its creators claim it does not strictly adhere to either misuse or anomaly detection paradigms, it tends to align more closely with the misuse paradigm, albeit from a basic standpoint. However, its fundamental design diverges from signature-based IDS by adopting an event-driven schema. Leveraging complex policies and Zeek's state awareness, it can also function as an anomaly detection system. In contrast to Snort, Zeek exhibits high state awareness, capable of maintaining states across connection boundaries. This enables the modeling of attack patterns based on events occurring hours or even days apart.

## 5. Feature tool comparison

In this section we conduct comparative analysis of tools for network traffic feature extraction. Key features of the most popular tools are presented here. All of selected tools are open-source software, but some have a full version with advanced options that could be unlocked with a paid subscription (Snort requires subscription for an access to up to date rules for IDS). Table 1 includes the list of chosen tools and their basic characteristics.

**Table 1**
Tools comparison

| Name of the tool | Level | Able to analyze real-time traffic |
|---|---|---|
| CICFlowmeter | Flow | No |
| Wireshark | Packet | Yes |
| Argus | Flow | Yes |
| Snort | Packet, application | Yes |
| Zeek | Packet, flow, application | Yes |

Wireshark is primarily a network protocol analyzer used for analyzing and troubleshooting network traffic. While it's not a dedicated Intrusion Detection System (IDS) like Zeek or Snort, it can be used as a component within an IDS setup. Wireshark captures packets on a network interface, making it possible to inspect the traffic in detail. Wireshark provides detailed analysis of captured packets, including used protocols, packet contents, source and destination addresses, etc. Patterns or anomalies in network traffic that may indicate suspicious or malicious activity can be detected via Wireshark's packet filtering and search capabilities.

Snort operates as a traditional IDS/IPS, conducting deep packet inspection and then applying signatures to traffic to detect and potentially block attacks. Therefore, it doesn't seem suitable for integration with AI systems.

In contrast, open-source software Zeek positions itself not as an IDS but as a network monitor and traffic analyzer. Its main function is to focus on comprehensive traffic inspection for signs of suspicious activity. Zeek supports a broad spectrum of traffic analysis tasks, extending beyond security to include performance measurement and troubleshooting.

Zeek's primary role appears to be capturing traffic details for external analysis systems. It emphasizes collecting comprehensive traffic information, sometimes integrating custom protocol dissectors tailored to the environment's protocols. While there's a functional overlap among these tools, their core objectives and utilization scenarios differ.

In this paper, we focus on using flow-based extraction tools, as they reduce the amount of resources required for processing data

**Table 2**
Comparison of flow-based tools

| Feature | Argus | Zeek | CICFlowMeter |
|---|---|---|---|
| Focus | Primarily focuses on network flow data analysis. It captures and analyzes network traffic to generate flow records, which provide detailed | Focuses on high-level network protocol analysis. It performs deep packet inspection to extract higher-level network | Primarily designed for flow-based network traffic analysis, particularly for |
| | insights into network activity. | events and metadata, providing detailed information about network activity, protocols, and behavior. | cybersecurity purposes, with a focus on detecting network intrusions, attacks, and anomalies. |
| Data Collection | Collects flow data, including information such as source and destination IP addresses, ports, protocols, and timestamps. | Collects detailed protocol-level metadata from network traffic, including HTTP, DNS, FTP, SMTP, and more. It can extract information such as HTTP headers, DNS queries, and file transfers, as well as flow data. | It collects flow data similar to Argus but with a specific focus on cybersecurity-related features such as attack detection and anomaly detection. |
| Analysis | Can be used to analyze network flow data to identify patterns, anomalies, and potential security threats. It's well-suited for analyzing traffic volume, trends, and basic behavior. | Performs deep protocol analysis to generate rich network logs. It can be used to detect complex network behaviors, such as reconnaissance activities, malware communication, and suspicious network traffic patterns. | Specializes in cybersecurity analysis, leveraging flow data to extract features from network data. |
| Flexibility | Offers flexibility in terms of capturing and exporting flow data, but its focus is primarily on flow analysis. | Highly extensible through its scripting language. Users can customize and extend its functionality to suit specific network monitoring and | Only available as Java tool used for feature extraction from network captures. |
| Community and ecosystem | Has a smaller user community compared to Zeek. It's widely used in certain sectors, particularly in academic and research environments. | Has a large and active user community, with a wealth of community-contributed scripts, plugins, and integrations. It's widely adopted across various industries, including cybersecurity, network operations, and research. | Used in some of the most popular IDS benchmark datasets. Beside that doesn't seem to be adopted anywhere. Also some researches criticize it. |

The following table includes the comparison of efficiency of the Random Forest classification model based on the same CIC-IDS2017 dataset with differently extracted features by CICFlowmeter and Zeek. The binary classification has been carried out, which labels network traffic as either "benign" (normal traffic) or "attack".

**Table 3**
Comparative results of Zeek and CICFlowmeter on CIC-IDS2017 [17]

| Name of the tool | LABEL | Random forest F1 score |
|---|---|---|
| CICFlowmeter | Benign | 0.994 |
| | Attack | 0.976 |
| Zeek | Benign | 0.998 |
| | Attack | 0.992 |

In [18], results showed a constant superiority of Netflow compared to CICFlowmeter. Moreover, according to [19, 20], the CICFlowMeter tool may present some incorrect implementation aspects both in the construction of the TCP protocol flows and in the extraction of attributes. Although the CICFlowTool was used in the most popular IDS benchmarking dataset (CIC-IDS-2017), it falls behind other tools compared here.

Because of all the criticism of CICFlowMeter presented, the most promising systems for IDS are Argus and Zeek.

In summary, while Argus and Zeek are valuable network monitoring tools, they have different strengths and are suitable for different use cases. Argus is ideal for flow data analysis and basic network traffic monitoring, while Zeek excels in deep protocol analysis and customizable network security monitoring.

## 6. Conclusions

Effective network traffic data analysis is essential for understanding, managing, and securing modern computer networks. It provides valuable insights that empower organizations to optimize performance, detect threats, and make informed decisions. As technology evolves, network traffic data analysis remains critical for network security.

The importance of AI-based IDS has been discussed here. They can be used from intrusion detection to user behavior analysis. Advancements in machine learning and big data analytics can enhance its capabilities. However, machine learning algorithms can not work with raw network flow, so feature space has to be constructed in advance. The comparative analysis of key tools that allow feature extraction for various applications has been conducted in this paper. Additionally, some flow-based tools for network feature extraction have been compared, highlighting the strengths and weaknesses of each tool. It was shown that the final efficiency of models depends on feature space and can differ even for the same raw network traffic. In the future, we plan to make a more extensive comparison of tools, including packet-based systems as well.